\documentclass[conference]{IEEEtran}
\IEEEoverridecommandlockouts
\usepackage{cite}
\usepackage{amsmath,amssymb,amsfonts}
\usepackage{graphicx}
\usepackage{textcomp}
\usepackage{xcolor}
\usepackage{url}
\usepackage{color}
\definecolor{bri}{rgb}{0.96, 0.73, 1.0}
\usepackage{colortbl}
\usepackage{lipsum}
\usepackage[inline]{enumitem}
\usepackage{balance}
\usepackage{tikz}
\usepackage{pgfplots}
\usepackage{algorithm}
\usepackage{algpseudocode}
\usepackage{verbatim}
\usepackage{float}

\DeclareRobustCommand{\hwplotA}{\raisebox{2pt}{\tikz{\draw[blue,solid,line width=0.9pt](0,0) -- (3mm,0);}}}
\DeclareRobustCommand{\hwplotB}{\raisebox{2pt}{\tikz{\draw[red,dashed,line width=1.2pt](0,0) -- (3mm,0);}}}
\DeclareRobustCommand{\hwplotC}{\raisebox{2pt}{\tikz{\draw[teal,thick,dotted,line width=1.2pt](0,0) -- (2.6mm,0);}}}

\usepackage{xcolor}

\usepackage{hyperref}

\def\BibTeX{{\rm B\kern-.05em{\sc i\kern-.025em b}\kern-.08em
    T\kern-.1667em\lower.7ex\hbox{E}\kern-.125emX}}
\begin{document}

\title{Pre-processing Blood-Volume-Pulse \\for In-the-wild Applications}

\author{\IEEEauthorblockN{1\textsuperscript{st} Laurits Fromberg}
\IEEEauthorblockA{\textit{Dept. Applied Mathematics and Computer Science} \\
\textit{Technical University of Denmark}\\
Kongens Lyngby, Denmark \\
s174512@student.dtu.dk}\\
\IEEEauthorblockN{3\textsuperscript{rd} Line Katrine Harder Clemmensen}
\IEEEauthorblockA{\textit{Dept. Applied Mathematics and Computer Science} \\
\textit{Technical University of Denmark}\\
Kongens Lyngby, Denmark  \\
lkhc@dtu.dk}
\and
\IEEEauthorblockN{2\textsuperscript{nd} Sneha Das}
\IEEEauthorblockA{\textit{Dept. Applied Mathematics and Computer Science} \\
\textit{Technical University of Denmark}\\
Kongens Lyngby, Denmark \\
sned@dtu.dk
}
}

\maketitle

\begin{abstract}
Blood-volume-pulse (BVP) is a biosignal commonly used in applications for non-invasive affect recognition and wearable technology.  However, its predisposition to noise constitutes limitations for its application in real-life settings. This paper re-visits BVP processing and proposes standard practices for feature extraction from empirical observations of BVP. We propose a method for improving the use of features in the presence of noise and compare it to a standard signal processing approach of a 4$^{\text{th}}$ order Butterworth bandpass filter with cut-off frequencies of 1~Hz and 8~Hz. Our method achieves better results for most time features as well as for a subset of the frequency features.
We find that all but one time feature and around half of the frequency features perform better when the noisy parts are known (best case). When the noisy parts are unknown and estimated using a metric of skewness, the proposed method in general works better or similar to the Butterworth bandpass filter, but both methods also fail for a subset features. Our results can be used to select BVP features that are meaningful under different SNR conditions.
\end{abstract}
\begin{IEEEkeywords}
Biosignals, Blood-Volume-Pulse,  In-the-wild, Feature Extraction, Pre-processing
\end{IEEEkeywords}

\section{Introduction}
Wearable technology has in recent years had a surge in popularity and attention from both customers as well as the scientific community. 
The area of applicability is extensive, embracing 
several aspects of everyday items.
A well-known case of modern wearable technology is the smart watch, which enables the acquisition of e.g., heart-rate, acceloremeter, and {GPS} leading to the analysis of traits such as physical activation, step counts, sleep patterns, etc. or even abstractions like \emph{body battery} \cite{wearable}.
One of the advantages of wearable devices is the continuous stream of personal data throughout all aspects of everyday life, even allowing for the acquisition of data during sleep.  
Wearable technology encompasses far-reaching possibilities within most areas of research. One of the most common being the use of biosignals for monitoring different aspects of ones personal health. 
However, wearable technology also yields itself to various challenges regarding applicability. This includes the quality and sensitivity of the data acquisition procedure being greatly influenced by 
external aspect \cite{wearable}. Another vital issue is the privacy threat associated with wearable technology due the sensitive nature of personal data \cite{wearable}. The applications and challenges of wearable technology thus constitute a wide variety of aspects.

This paper addresses the challenges relating to 
noise in observations of blood-volume-pulse (BVP). BVP is a biosignal often applied in the setting of non-invasive emotion recognition with the use of machine learning algorithms. However, it is to a great extent prone to noise especially in comparison to other biosignals,
hence limiting its usability in real-life settings.
The reasons for this include
motion, lack of tightness of measurement device causing inconsistent illumination, participant dependent characteristics, etc. \cite{b1}. These uncertainties and shortcomings cause difficulties for fully exploiting the advantages of the signal.
This is a well-known issue, which has been widely discussed within the field of research \cite{b1}\cite{b2}\cite{b4}, 
hence why it is of interest to assess good practices, in order to ease use of applications and analyses exploiting the biosignal within settings requiring feature extraction such as it is often the case in for instance emotion recognition.

\subsection{Related Work}
Emotion recognition using non-invasive wearable devices such as for instance the Empatica E4 \cite{b4} has in recent years been of great interest in the field of research, where machine learning models have been used to establish stress-recognition systems \cite{stress}, forecasting seizures \cite{seizure}, etc. 
However, due to the great variation within individuals as well as the presence of noise in real-life settings predictions often lack robustness, hence having difficulties generalising outside the constraint of ambulatory settings \cite{b9}.
This is in particular an issue for the BVP biosignal which is prone to noise, hence limiting its applicability to real-life settings \cite{b1}\cite{b10}.
Various signal-processing approaches have been proposed in the literature for extracting relevant information from the otherwise inherently noisy signal \cite{bvp_filt}\cite{b9}, although there appears to be a gap in the research in regards to addressing the influence of noise on specific features as well as establishing a framework of good practices for utilising noisy BVP signals.
Hence, the main contributions of this paper are the following:\\
\begin{enumerate}
    \item Establish good practices for feature extraction of BVP in the setting of noise based upon empirical observations.\\
    \item Propose a method for extracting features from a BVP signal in the presence of noise. 
\end{enumerate}

\section{Methodology}

\subsection{Features}

For the assessment of good practices for BVP we consult various features. The features include both linear and non-linear features in the time domain as well as the frequency domain in order to embrace a comprehensive overview. A description is available in table \ref{features}. Note that features in the frequency domain were separated into real and imaginary parts. 
\begin{table}[!htbp]
    \centering
    \renewcommand{\arraystretch}{1.2}
    \caption{Feature Descriptions.}
    \begin{tabular}{ccc}
  Domain & Feature & Description \\
  \hline
  \hline
\textbf{Both} & \textbf{Avg.} & Average  \\
\hline
\textbf{Both} & \textbf{Std.} & Standard deviation  \\
\hline
\textbf{Both} & \textbf{Med.} & Median  \\
\hline
\textbf{Both} & \textbf{Min.} & Minimum  \\
\hline
\textbf{Both} & \textbf{Max.} & Maximum \\
\hline
\textbf{Time} & \textbf{Int.} & Integral (trapezoid-rule) \\
\hline
\textbf{Time} & \textbf{Min. S.} & Minimum slope  (sliding window) \\
\hline

\textbf{Time} & \textbf{Max. S.} & Maximum slope (sliding window) \\
\hline
\textbf{Time} & \textbf{Avg. S.} & Average slope (sliding window) \\
\hline
\textbf{Time} & \textbf{Slope} & Overall slope  \\
\hline
\textbf{Time} & \textbf{Avg. G.} & Average gradient (finite-difference)\\
\hline
\textbf{Time} & \textbf{Avg. N. G.} & Average negative gradient\\
\hline
\textbf{Frequency} & \textbf{Sum.} & Sum of frequencies\\
\hline
\textbf{Frequency} & \textbf{IQR.} & Interquantile Range\\
\hline
\hline
    \end{tabular}
    \label{features}
\end{table}

\subsection{Experimental Design}
\begin{algorithm}
 \begin{algorithmic}
  \caption{Algorithm for generating noisy signal}\label{alg:noise}
\State $L \gets 0.5*F_s$\Comment{Length of noise segment}
\State $S\gets round\{275*Fs/L\}$ \Comment{Number of noise increments}
\State $M \gets 50$ \Comment{Number of Iterations}
\While{$m < M$}
\State $y_{noisy} \gets x_{bvp}$\Comment{Clean BVP signal}
\While{$s < S$}
    \State $y_{noisy}[Ls:L(s+1)] \gets n_{white}$ \Comment{White-noise substitution}
        \State $y_{noisy}[r_1:r_1+L] \gets y_{noisy}[r_1:r_1+L] + n_{HF}$ \Comment{Add high-frequency noise}
\State $j_{impulse} \gets r_2$
     \If{$j_{impulse} = 1$}
       \State $y_{noisy}[r_3] \gets y[r_3]\pm n_{impulse} $\Comment{Add noise-impulse}
    \EndIf
    \EndWhile
    \EndWhile
 \end{algorithmic}
\end{algorithm}

To retrieve as much information as possible from the noisy BVP, we develop an experimental design reflecting the problem at hand. 
We simulate a BVP signal with a heart-rate of 70 beats-per-minute, a sample-rate of 64Hz, and a duration of 300 seconds and extract the features described in the previous section using the implementation made available in \cite{b8}.
We then add various sources of noise to the signal in an incremental fashion, that is; we achieve three different versions of the same signal identified by the addition of different noise sources. The process for generating the noisy signal is illustrated in Alg.~\ref{alg:noise} and is described below:
\begin{enumerate}
    \item Substitute white-noise in 0.5 second increments from the start of the signal, with 550 increments in total. 
    \item Add segments of high-frequency noise to random positions of version 1. The noise has a lower cut-off at 20Hz.
    \item Add high-intensity impulses (with probability $p=0.1$) to random positions of version 2.
\end{enumerate}
The selection of the above noise sources have been done in accordance with the noise present in a BVP signal acquired during a pilot-study experiment to achieve a thorough representation. We then extract the features from the noisy signals and compare these with the features from the original signal.

\subsection{Baseline method}
We compare our method to a standard signal-processing approach for BVP signals; a $4^{\text{th}}$ order Butterworth bandpass filter with cut-off frequencies of 1Hz and 8Hz \cite{b9}. This is done for the signal after step 3 of the noise adding procedure and without removing any parts of the signal. 
\subsection{Segmentation Method}
\begin{algorithm}
 \begin{algorithmic}
  \caption{Segmentation algorithm}\label{alg:segmentation}
  \State $L\gets len{x}$
\State $W_L \gets 5*F_s$\Comment{Window-length}
\State $W_S \gets 1*F_s$\Comment{Window-shift}
\State $T \gets 0.05$\Comment{Skew-threshold}
\State $ind \gets 1:L$
\While{$i<L$}\Comment{Segmenting the signal}
    \State $Win = append(ind[i:(i+W_L)])$
    \State $i = i+W_L-W_S$
    \EndWhile
\While{$j<len(Win)$} \Comment{Goodness of fit}
\State $p = Skew(Win[j])$
\If {$p>T$}
\State $Segs \gets append(Win[j])$
\EndIf
\EndWhile

\State $x_{filt} \gets bandpass(Segs)$ \Comment{Filtering valid segments}
\While {$k<num(x_{filt})$}
\State $X[k] \gets feature(x_{filt}[k])$\Comment{Extract features}
\EndWhile
\State $X_{output}\gets \frac{1}{K}\sum\limits_{k}X[k]$ 
 \end{algorithmic}
\end{algorithm}

We propose to remove the parts of the BVP signal that are influenced by noise and compute the features of each remaining segment and then taking an average of the computed features. We refer to this procedure as the \textit{segmentation method}.
Our proposed algorithm, demonstrated in Alg.~\ref{alg:segmentation} is as follows:
\begin{enumerate*}
    \item Use a sliding window with suitable specifications of the window-size and window-shift in order to extract segments from the signal (We use 5 seconds and 1 second respectively).
    \item Assess which segments are appropriate according to a measure of fit. We propose to use the skewness as a valid index of quality for discriminating between different standards of BVP as described in \cite{b100}. Keep the segments the have skewness $<0.05$ and exclude the rest.
    \item Ensure that each segment has more points than a pre-specified threshold. Discard the segments, which do not fulfill this requirement. 
    \item Apply a 2$^{\text{nd}}$ order Butterworth bandpass filter with cut-off frequencies 1~Hz and 8~Hz, to remove noise from the remaining segments.
    \item Compute the features of each remaining segment and output their average, weighted by the number of points in each of the corresponding segments.
\end{enumerate*}
As a best-case baseline, we will test the results of this algorithm if we simply remove noisy parts of the signal instead of estimating which parts are noisy.

\subsection{Varying Segment-Size} Another experiment was established in order to assess a suitable segment-size. A larger segment-size could conceivably enable the acquisition of more information at lower frequencies, whereas a smaller segment-size would yield an adequate representation at larger frequencies. We varied the segment-size and inspected the corresponding differences between the features of the original signal and the features of the noisy signal derived using the segmentation method.

\section{Results and Discussion}
\begin{figure*}[htbp]
    \centering
    \includegraphics[width=\textwidth]{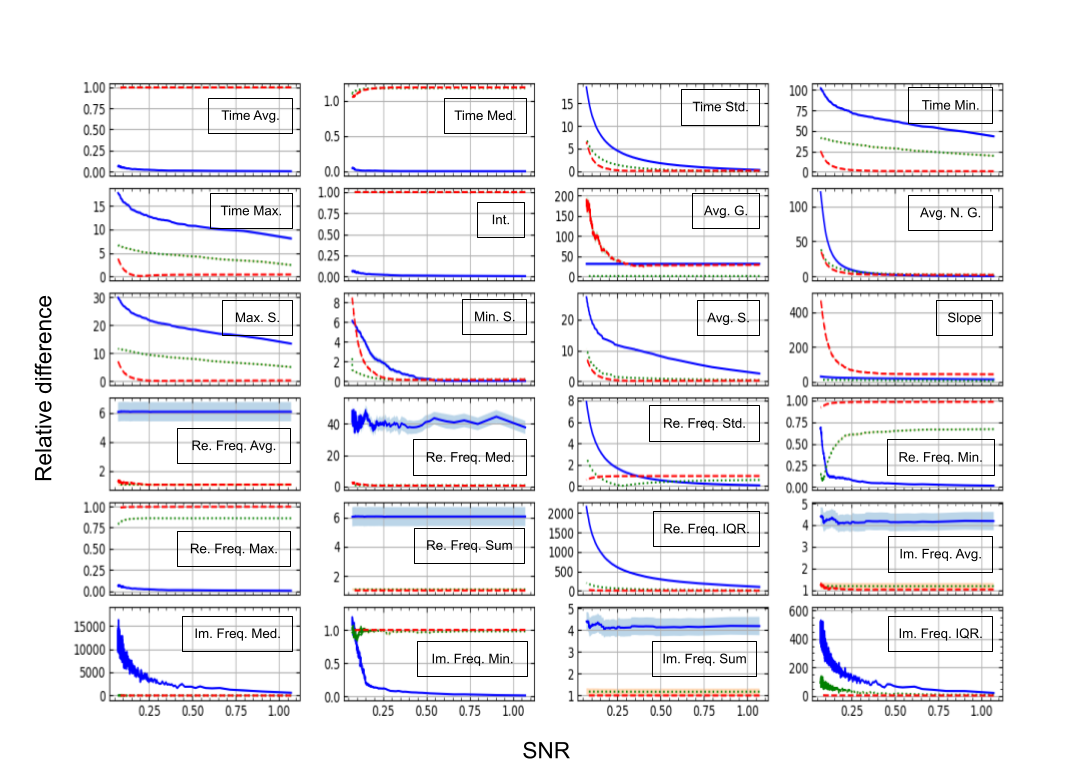}
       \caption{Differences between extracted BVP features and the true feature (from BVP signals without noise) as a function of SNR. BVP features are extracted from:
 raw noisy signal [{\hwplotA}],
 $4^{\text{th}}$  order Butterworth bandpass filter (1Hz-8Hz) [{\hwplotC}],
segmentation method [{\hwplotB}]. The average of 50 iterations $\pm1$ standard error of the mean (shaded region).}
    \label{fig:bvp_segmentation}
\end{figure*}

We start by visualising the result of the aforementioned experiment in figure \ref{fig:bvp_segmentation} by plotting the relative difference between the original features to the noise altered features as a function of the signal-to-noise-ratio i.e. $\text{SNR}=\mu/\sigma$ \cite{b3}. Here, we use a window-size of 5 seconds and a window-shift of 1 seconds for the sliding window.
Furthermore, the results and analysis in relation to the segmentation method are presented in figure \ref{fig:bvp_segmentation}. 

In order to assess if the segmentation method attains better results on average than the standard signal-processing approach, we perform dependent two-sample $t$-tests. The results of which can be found in table \ref{tab:t_test}. Performing $88$ hypothesis tests yields for the pre-specified significance level of $\alpha=0.05$, a family-wise error rate of approximately $0.989$.
Therefore, to prevent the problem of multiple hypothesis testing we adopt the Bonferroni correction. This yields a corrected significance level of approximately $0.0006$.
 \begin{table}[htbp]
    \caption{The one-sided right-tailed $t$-tests for testing if the segmentation method is significantly better than the standard signal-processing approach on average. Here presenting the respective $p$-values. Statistical significance is shown by highlighting the appropriate cell.}
    \centering
  
    \begin{tabular}{|l|c|c|c|c|}
    \hline
       \text{Feature/SNR}  & 0.10 & 0.25 & 0.50 & 0.75\\ \hline
       \hline
       \text{Time Avg.}   &  \cellcolor{bri}0.0000 &   \cellcolor{bri}0.0000 &   \cellcolor{bri}0.0000 &   \cellcolor{bri}0.0000 \\\hline
\text{Time Med.}      &   \cellcolor{bri}0.0000 &   \cellcolor{bri}0.0000 &   \cellcolor{bri}0.0000 &   \cellcolor{bri}0.0000 \\\hline
\text{Time Std.}        &   \cellcolor{bri}0.0000 &   \cellcolor{bri}0.0000 &   \cellcolor{bri}0.0000 &   \cellcolor{bri}0.0000 \\\hline
\text{Time Min.}        &   \cellcolor{bri}0.0000 &   \cellcolor{bri}0.0000 &   \cellcolor{bri}0.0000 &   \cellcolor{bri}0.0000 \\\hline
\text{Time Max.}         &   \cellcolor{bri}0.0000 &   \cellcolor{bri}0.0000 &   \cellcolor{bri}0.0000 &   \cellcolor{bri}0.0000 \\\hline
\text{Int.}     &   \cellcolor{bri}0.0000 &   \cellcolor{bri}0.0000 &   \cellcolor{bri}0.0000 &   \cellcolor{bri}0.0000 \\
\hline
\text{Max. S.} &   \cellcolor{bri}0.0000 &   \cellcolor{bri}0.0000 &   \cellcolor{bri}0.0000 &   \cellcolor{bri}0.0000 \\
\hline
\text{Min S.}       &   \cellcolor{bri}0.0000 &   \cellcolor{bri}0.0010 &   \cellcolor{bri}0.0000 &   \cellcolor{bri}0.0000 \\\hline
\text{Avg. S.}       &   \cellcolor{bri}0.0000 &   \cellcolor{bri}0.0000 &   \cellcolor{bri}0.0000 &   \cellcolor{bri}0.0000 \\\hline
\text{Slope}    &  1.0000 &  1.0000 &  1.0000 &  0.9998 \\
\hline
\text{Re. Freq. Avg.}   &  0.0972 &  0.2210 &  0.2021 &  0.5422 \\\hline
\text{Re. Freq. Med.} &  0.9977 &  1.0000 &  0.9833 &  0.9777 \\\hline
\text{Re. Freq. Std.}   &  \cellcolor{bri} 0.0000 &  1.0000 &  1.0000 &  1.0000 \\\hline
\text{Re. Freq. Min.}    &  1.0000 &  1.0000 &  1.0000 &  1.0000 \\\hline
\text{Re. Freq. Max.}    &  1.0000 &  1.0000 &  1.0000 &  \cellcolor{bri}0.0000 \\\hline
\text{Re. Freq. Sum}    &  0.0009 &  \cellcolor{bri}0.0005 &   \cellcolor{bri}0.0000 &   \cellcolor{bri}0.0000 \\\hline
\text{Re. Freq. IQR.}    &  \cellcolor{bri}0.0000 &   \cellcolor{bri}0.0000 &   \cellcolor{bri}0.0000 &  \cellcolor{bri} 0.0000 \\\hline
\text{Im. Freq. Avg.}    &   \cellcolor{bri}0.0000 &   \cellcolor{bri}0.0000 &  \cellcolor{bri} 0.0000 &   \cellcolor{bri}0.0000 \\\hline
\text{Im. Freq. Med.}   &  0.1238 &  0.0363 &  0.0460 &  0.0069 \\\hline
\text{Im. Freq. Min.}    &  0.0579 &  0.0401 &  0.0543 &  0.7305 \\\hline
\text{Im. Freq. Sum}     &   \cellcolor{bri}0.0000 &   \cellcolor{bri}0.0000 &   \cellcolor{bri}0.0000 &   \cellcolor{bri}0.0000 \\\hline
\text{Im. Freq. IQR.}    &   \cellcolor{bri}0.0000 &   \cellcolor{bri}0.0000 &   \cellcolor{bri}0.0000 &  \cellcolor{bri} 0.0000 \\\hline
\end{tabular}

\label{tab:t_test}
 \end{table}

The study in relation to the BVP experiment demonstrates, as seen in figure \ref{fig:bvp_segmentation} that; 
\begin{enumerate}
    \item The signal-to-noise-ratio appears to have immense influence on the quality of the features. 
    \item For well-behaved signals, i.e. $\text{SNR}>1$, all features except the sum of frequencies, the average frequency, the real part of the median frequency, the time minimum, the time maximum and the average gradient are suitable for emotion recognition. While, the maximum slope and the slope ought to be the subject of great scrutiny as the established errors associated vary in magnitude.
    \item For acceptable signals, i.e. $\text{SNR}\in [0.5,1]$, a subset of features are still sensible for emotion recognition, albeit the need for reluctance in regards 
    to the features of the minimum slope, the average slope as well as the interquantile range of the frequencies.
    \item For substandard signals, i.e. $\text{SNR}<0.5$, most features are not reasonable to exploit and one should thus consider discarding the signal. However, if necessary one may (with caution) consider using the features; the time average, the time median, the integral, the minimum frequency, the maximum frequency as well as the average negative gradient. 
\end{enumerate}
These observations constitute the circumstances and criteria for which the different features of BVP are appropriate for emotion recognition. Note that one should be reluctant to conclude upon these circumstances and criteria from version 1 of the noise altering procedure, as this version may not depict real-life noise sources adequately, whereby a more "worst-case" approach is deemed necessary, and the reason for considering version 3.
Furthermore, it must be noted that the most important characteristics of the BVP signal is the frequency and the morphology, and not the intensity \cite{b4}. In relation hereto, on account of the aforementioned observations, that the time features may perhaps appear to have a predisposition to robustness of noise altercations, in contrast to the frequency counterparts.
It is likewise seen in relation to the performance of the segmentation method from figure \ref{fig:bvp_segmentation} that;
\begin{itemize}
    \item In general, the segmentation method appears to produce an improvement in the results, except for the features; the integral, the slope, the minimum of frequencies, the average gradient and the maximum of frequencies. 
    \item The segmentation method yields similar performance to the standard signal-processing approach. 
\end{itemize}
The reason for the poor performance of the features in question ought to be on account of
the signal being discontinuous. Thus 
making the features comprise a linear combination of features originating from each individual segment, under which condition it is not necessarily sensible to evaluate the aforementioned features. 
Hereby also changing the circumstances and basis for which we compare the features, as these features now arise differently from the original features which they are indeed compared to, thus yielding a discrepancy. Ergo asserting that the method is not applicable for these specific features, to prevent the attainment of unreasonable results which may otherwise be the case.

Additionally consulting table \ref{tab:t_test} it is seen that the segmentation methods on average performs significantly better than the classical signal-processing approach for a subset of the features at the specific values of the signal-to-noise-ratio. 
The specific features, which yield better performance with the segmentation method consist of all the time features except the slope as well as a subset 
of the frequency features including; the interquantile range of the frequencies, the real part of the sum of frequencies (except for $\text{SNR}=0.10$), the imaginary part of the sum of frequencies  as well as the imaginary part of the average of the frequencies.

The segmentation method is thus seen to provide promising results with the method in some cases
outperforming the standard signal-processing approach, as seen from table \ref{tab:t_test}. 
The observations from the segmentation method thus facilitates good practices for BVP in the setting of machine learning models where feature extraction constitutes a necessity. 
In regards to the experiment inspecting the use of varying the segment-size it was found that it was not possible to establish a satisfactory segment-size, as the results did not provide grounds for a general conclusion.

\section{Conclusions and Future Work}
\label{sec:con&fut}
In this paper, we explore a framework for using the BVP under low signal-to-noise ratio with respect to feature extraction. While there is no clear demarcation on the features that perform well under the constraint of noise, empirical observations indicate the applicability of a subset of time and frequency features, as these yield adequate performance under different thresholds settings of signal-to-noise-ratios.
Summing up, we were able to devise a method for improving the use of features in the presence of noise, which in many cases yield significantly better results than a standard signal processing approach of a $4^{\text{th}}$ order Butterworth bandpass filter with cut-off frequencies of 1Hz and 8Hz. This improvement is seen for almost all the time features as well as a subset of frequency features.


As our best-case baseline has a better performance for substantially more features than with our skewness to determine noisy signal parts, future work will be devoted to other measures for deciding when a signal is so noisy that it needs removed. 






\begin{thebibliography}{00}
\bibitem{b1} P. J. Bota, C. Wang, A. L. N. Fred and H. Plácido Da Silva, "A Review, Current Challenges, and Future Possibilities on Emotion Recognition Using Machine Learning and Physiological Signals," in IEEE Access, vol. 7, pp. 140990-141020, 2019, doi: \href{10.1109/ACCESS.2019.2944001}{https://ieeexplore.ieee.org/document/8849996}.

\bibitem{b2} Luo, Simon \& Zhou, Jianlong \& Duh, Henry \& Chen, Fang. (2017). BVP Feature Signal Analysis for Intelligent User Interface. 1861-1868. 10.1145/3027063.3053121. 

\bibitem{b3} Elkum, Naser \& Shoukri, Mohamed. (2008). Signal-to-noise ratio (SNR) as a measure of reproducibility: Design, estimation, and application. Health Services and Outcomes Research Methodology. 8. 119-133. 10.1007/s10742-008-0030-2. 

\bibitem{b4} The Empatica E4 Research. \url{https://www.empatica.com/en-int/research/e4/}.
\bibitem{stress} Chen, J.; Abbod, M.; Shieh, J.-S. Pain and Stress Detection Using Wearable Sensors and Devices—A Review. Sensors 2021, 21, 1030. \url{https://doi.org/10.3390/s21041030}.

\bibitem{seizure} Nasseri, M., Pal Attia, T., Joseph, B. et al. Ambulatory seizure forecasting with a wrist-worn device using long-short term memory deep learning. Sci Rep 11, 21935 (2021). \url{https://doi.org/10.1038/s41598-021-01449-2}.

\bibitem{wearable} 
Aleksandr Ometov, Viktoriia Shubina, Lucie Klus, Justyna Skibińska, Salwa Saafi, Pavel Pascacio, Laura Flueratoru, Darwin Quezada Gaibor, Nadezhda Chukhno, Olga Chukhno, Asad Ali, Asma Channa, Ekaterina Svertoka, Waleed Bin Qaim, Raúl Casanova-Marqués, Sylvia Holcer, Joaquín Torres-Sospedra, Sven Casteleyn, Giuseppe Ruggeri, Giuseppe Araniti, Radim Burget, Jiri Hosek, Elena Simona Lohan,
A Survey on Wearable Technology: History, State-of-the-Art and Current Challenges,
Computer Networks,
Volume 193,
2021,
108074,
ISSN 1389-1286,
\href{https://www.sciencedirect.com/science/article/pii/S1389128621001651}{https://doi.org/10.1016/j.comnet.2021.108074}



\bibitem{bvp_filt} Sabeti, E.; Gryak, J.; Derksen, H.; Biwer, C.; Ansari, S.; Isenstein, H.; Kratz, A.; Najarian, K. Learning Using Concave and Convex Kernels: Applications in Predicting Quality of Sleep and Level of Fatigue in Fibromyalgia. Entropy 2019, 21, 442. \url{https://doi.org/10.3390/e21050442}.

\bibitem{b10} Saganowski, S. Bringing Emotion Recognition Out of the Lab into Real Life: Recent Advances in Sensors and Machine Learning. Electronics 2022, 11, 496. \url{https://doi.org/10.3390/electronics11030496}.



\bibitem{b8} Makowski, D., Pham, T., Lau, Z. J., Brammer, J. C., Lespinasse, F., Pham, H., Schölzel, C., \& Chen, S. A. (2021). NeuroKit2: A Python toolbox for neurophysiological signal processing. Behavior Research Methods, 53(4), 1689-1696. \url{https://doi.org/10.3758/s13428-020-01516-y}.

\bibitem{b9}Yaqian Xu, I. Hübener, A. -K. Seipp, S. Ohly and K. David, "From the lab to the real-world: An investigation on the influence of human movement on Emotion Recognition using physiological signals," 2017 IEEE International Conference on Pervasive Computing and Communications Workshops (PerCom Workshops), Kona, HI, USA, 2017, pp. 345-350, doi: \href{https://ieeexplore.ieee.org/document/7917586/}{10.1109/PERCOMW.2017.7917586}.

\bibitem{b100} Elgendi M. Optimal Signal Quality Index for Photoplethysmogram Signals. Bioengineering (Basel). 2016 Sep 22;3(4):21. doi:\href{https://www.mdpi.com/2306-5354/3/4/21}{10.3390/bioengineering3040021}. PMID: 28952584; PMCID: PMC5597264.
\end{thebibliography}
\end{document}